\documentclass[traditabstract]{aa}
\usepackage{subfigure}
\usepackage{longtable}
\usepackage{graphicx}
\usepackage{graphics}
\usepackage{keyval}
\usepackage{trig}
\usepackage{psfig,epsf}
\usepackage{txfonts}
\def\beq{\begin{equation}}
\def\eeq{\end{equation}}

\begin{document}

\title{THINGS about MOND}

\author{G. Gentile\inst{1} \and B. Famaey\inst{2,3} \and W. J. G. de Blok\inst{4}}

\institute{Sterrenkundig Observatorium, Universiteit Gent, Krijgslaan 281, B-9000 Gent, Belgium\\
              \email{gianfranco.gentile@ugent.be}
         \and
          Observatoire Astronomique,  Universit\'e de Strasbourg, CNRS UMR 7550, 
F-67000 Strasbourg, France
         \and
         AIfA, Universt\"at Bonn, D-53121 Bonn, Germany
         \and
         Department of Astronomy, University of Cape Town, Private Bag X3, Rondebosch 7701, South Africa
}

\abstract
{We present the analysis of 12 high-resolution galactic rotation curves from The HI Nearby Galaxy Survey (THINGS) in the context of modified Newtonian dynamics (MOND).
These rotation curves were selected to be the most reliable for mass modelling, and they are the highest quality rotation curves currently
available for a sample of galaxies spanning a wide range of luminosities.
We fit the rotation curves with the ``simple'' and ``standard" interpolating functions of MOND, and we find 
that the ``simple" function yields better results.
We also redetermine the value of $a_0$, and find a median value very close to the one determined in previous studies, $a_0$ = (1.22 $\pm$ 0.33) $\times$ 10$^{-8}$ cm s$^{-2}$.
Leaving the distance as a free parameter
within the uncertainty of its best independently determined value leads to excellent quality fits for 75\% of the sample. Among the
three exceptions, two are also known to give relatively poor fits also in Newtonian dynamics plus dark matter.
The remaining case (NGC 3198), presents some tension between the observations and the MOND fit, 
which might however be explained by the presence of non-circular motions, by a small distance, or by a value of $a_0$ at the lower end of our best-fit interval, 0.9 $\times$ 10$^{-8}$ cm s$^{-2}$.
The best-fit stellar $M/L$ ratios are generally in remarkable agreement with the predictions of
stellar population synthesis models. We also show that the narrow range of gravitational accelerations found to be generated by dark matter in galaxies is consistent with the narrow range of additional gravity predicted by MOND.}

\keywords{galaxies: kinematics and dynamics - dark matter - galaxies: spiral -
gravitation}

\maketitle

\section{Introduction}
\protect\label{sec:intr}
The current dominant paradigm is that galaxies are embedded in halos of cold dark matter (CDM), made of non-baryonic weakly-interacting massive particles (e.g., Bertone et al. 2005). However, an alternative way to explain the observed rotation curves of galaxies is the postulate of Milgrom (1983) that for gravitational accelerations below a certain value $a_0$, the true gravitational attraction $g$ approaches $(g_N a_0)^{1/2}$ where $g_N$ is the usual Newtonian gravitational field (as calculated from the observed distribution of visible matter): this paradigm is known as modified Newtonian dynamics (MOND).

MOND explains successfully many phenomena in galaxies, among which the following non-exhaustive list: (i) it predicted the shape of rotation curves of low surface-brightness (LSB) galaxies before any of them had ever been measured (e.g. McGaugh \& de Blok 1998); (ii) tidal dwarf galaxies (TDG), which should be devoid of collisionless dark matter, still exhibit a mass-discrepancy in Newtonian dynamics, which is perfectly explained by MOND (Gentile et al. 2007); (iii) the baryonic Tully-Fisher relation (e.g., McGaugh 2005), one of the tightest observed relations in astrophysics, is a natural consequence of MOND, both for its slope and its zero-point; (iv) the first realistic simulations of galaxy merging in MOND were recently carried out, notably reproducing the morphology of the Antennae galaxies (Tiret \& Combes 2008); (v) it naturally explains the 
universality of ``dark" and baryonic surface densities within one core radius in galaxies (Donato et al. 2009, Gentile et al. 2009). 

Recent theoretical developments have also added plausibility to the case for MOND through the work of, e.g., Bekenstein~(2004), Sanders~(2005), Zlosnik, Ferreira \& Starkman~(2007), Halle, Zhao \& Li~(2008), and Blanchet \& Le Tiec~(2008), who have all presented Lorentz-covariant theories yielding a MOND behavior in the weak field limit.  Although still fine-tuned and far from a fundamental theory explaining the MOND paradigm, these effective theories remarkably allow for new predictions regarding cosmology (e.g., Skordis et al. 2006) and gravitational lensing (e.g., Angus et al.~2007, Shan et al.~2008). For reviews of MOND's successes and weaknesses, both at the observational and theoretical level, as well as comparisons with dark matter results, see McGaugh \& de Blok~(1998), 
de Blok \& McGaugh (1998), Sanders \& McGaugh~(2002), Bruneton \& Esposito-Far\`ese~(2008), Milgrom~(2008), Skordis~(2009), Famaey \& Bruneton~(2009), Ferreira \& Starkman~(2009).

One thing the MOND paradigm does not directly predict, though, is the shape of the interpolation between the MONDian regime where $g \ll a_0$ and the Newtonian regime where $g \gg a_0$, as well as the actual value of the acceleration constant $a_0$. The latter is in principle a free parameter, but once its value has been determined by some means, it must be identical for every astronomical object. Large variations of $a_0$ would invalidate MOND as a fundamental paradigm underpinned by new physics. Let us note that, as shown in Begeman, Broeils \& Sanders~(1991) fits with variable $a_0$ and fixed distance $D$ are essentially identical to fits with fixed $a_0$ and variable $D$ because the observed total gravitational acceleration is proportional to
$1/D$. Ideally, the fitted distance should however generally conform to the independently determined one (e.g., Cepheids-based or RGB tip-based). Finally, a consequence of the absence of galactic dark matter within the MOND context is that the dynamical mass-to-light ratio that is derived from a rotation curve fit should agree with the true stellar mass-to-light ratio of the stellar disk (and sometimes bulge), as inferred from
e.g. observed colours and stellar population synthesis models. 

Here, we use results from The HI Nearby Galaxy Survey (THINGS; Walter et al 2008), which consists of high-resolution HI observations of a sample of 34 nearby galaxies, in order to constrain the transition function of MOND. In particular, we show that some individual galaxies that had been claimed to be potentially problematic for MOND such as NGC~2841 (Begeman et al. 1991) can yield good fits with the ``simple" interpolating function.

We use a subset of the THINGS galaxies for which rotation curves could be derived in de Blok et al.~(2008), restricting ourselves to galaxies which are not (obviously) dominated by non-circular motions. In Sect.~2, we summarize the popular choices that have been proposed in the literature for the transition between the MONDian and Newtonian regimes, in order to confront these different transitions with THINGS rotation curves. In Sect.~3, we explain how we selected the subsample of galaxies that we model in the context of MOND. Sect.~4.1. then presents the results for the value of the acceleration constant $a_0$, while Sect.~4.2. and 4.3. present the comparison of the rotation curve fits for the different transitions, especially the best-fit mass-to-light ratios and distances. Finally, in Sect.~4.4., we discuss NGC 3198, the only cases where the MOND fits perform significantly worse than dark matter fits in the context of Newtonian dynamics. Conclusions are drawn in Sect.~5.

\section{MOND and its interpolating function}
\protect\label{sec:mu}
The MOND paradigm stipulates that the Newtonian acceleration $\vec{g}_N$ produced by the visible matter is linked to the true gravitational acceleration $\vec{g}$ by means of an interpolating function $\mu$:
\begin{equation}
\mu\left(\frac{g}{a_{0}}\right)\vec{g} = \vec{g}_{N},
\label{eq:A}
\end{equation}
where
$\mu(x) \sim x$ for $x \ll 1$ and
$\mu(x) \sim 1$ for $x \gg 1$ (and $g=|\vec{g}|$). However, this expression cannot be exact for all orbits and all geometries, since it does not respect usual conservation laws. Such a modification of Newtonian dynamics could come at the classical (non-covariant) level from a modification of either the kinetic part or the gravitational part of the Newtonian action (with usual notations; $\phi_N$ being the Newtonian gravitational potential):
\begin{equation}
S=\int \frac{1}{2}\rho v^2 d^3x \, dt \, - \int \left(\rho \phi_N + \frac{|\nabla \phi_N|^2}{8 \pi G}\right) d^3x \, dt,
\end{equation}
where modifying the first term is referred to as {\it modified inertia} and  modifying the second term as {\it modified gravity}. Milgrom~(1994) has shown that within the modified inertia framework, Eq.~1 was exact {\it only} for circular orbits (for other orbits, predictions are difficult to make since the theory is non-local). On the other hand, Bekenstein \& Milgrom~(1984) have shown that within a modified gravity framework where $|\nabla \phi_N|^2$ is replaced by $a_0^2 F(|\nabla \phi|^2/a_0^2)$ in Eq.~2 ($\phi$ being the MONDian potential and $F'=\mu$), the right-hand side of Eq.~1 had to be replaced by $\vec{g}_N + \vec{s}$ where $\vec{s}$ is a solenoidal vector field determined by the condition that $\vec{g}$ can be expressed as the gradient of a MONDian potential. Milgrom~(2010) has proposed another modified gravity formulation in which $|\nabla \phi_N|^2$ is replaced by $2 \nabla \phi \cdot \nabla \phi_N - a_0^2 Q(|\nabla \phi_N|^2/a_0^2)$ ($\phi$ being the MOND potential, $\phi_N$ remaining the Newtonian one, and $1/Q'=\mu$): in this case, the solenoidal field to be added to the right-hand-side of Eq.~1 is different from the one in the Bekenstein \& Milgrom formulation (see also Zhao \& Famaey~2010).

Although Brada \& Milgrom~(1995), Famaey et al.~(2007) and Zhao \& Famaey (2010) have shown that the expected differences in the predictions of the various formulations for rotation curves are not very large, they can be of the same order of magnitude as the differences produced by different choices for the $\mu$-function. In order to constrain $\mu$ within the modified gravity framework, one should calculate predictions of the modified Poisson formulations of Bekenstein \& Milgrom~(1984) or Milgrom~(2010) numerically for each galaxy model, and for each choice of parameters. This is left for further works and we choose here to concentrate on the modified inertia formulation for circular orbits given by Eq.~1.

It is worth noting that other interesting constraints on MOND and its $\mu$-function could come from studies of the effect of the galactic gravitational field on the dynamics of the inner Solar System (Milgrom 2009), or from studies of the dynamics perpendicular to the galactic disk at the solar position (Bienaym\'e et al. 2009).

Various choices for the shape of the $\mu$-function have been proposed in the literature (see especially Milgrom \& Sanders~2008 and McGaugh~2008), but we rather concentrate here on the two most popular choices that have been studied so far. The ``standard" $\mu$-function (Milgrom 1983) yields a relatively sharp transition from the MONDian ($x \ll 1$,
where $x=g/a_0$ and $g$ is the gravitational acceleration) to the Newtonian ($x \gg 1$) regime:
\begin{equation}
\mu(x)= \frac{x}{\sqrt{1+x^2}},
\label{eqstandard}
\end{equation}
while the ``simple" $\mu$-function (Famaey \& Binney 2005; Zhao \& Famaey 2006) yields a more gradual transition:
\begin{equation}
\mu(x) =\frac{x}{1+x}.
\label{eqsimple}
\end{equation}

\begin{figure}
\begin{center}
\includegraphics[scale=0.4]{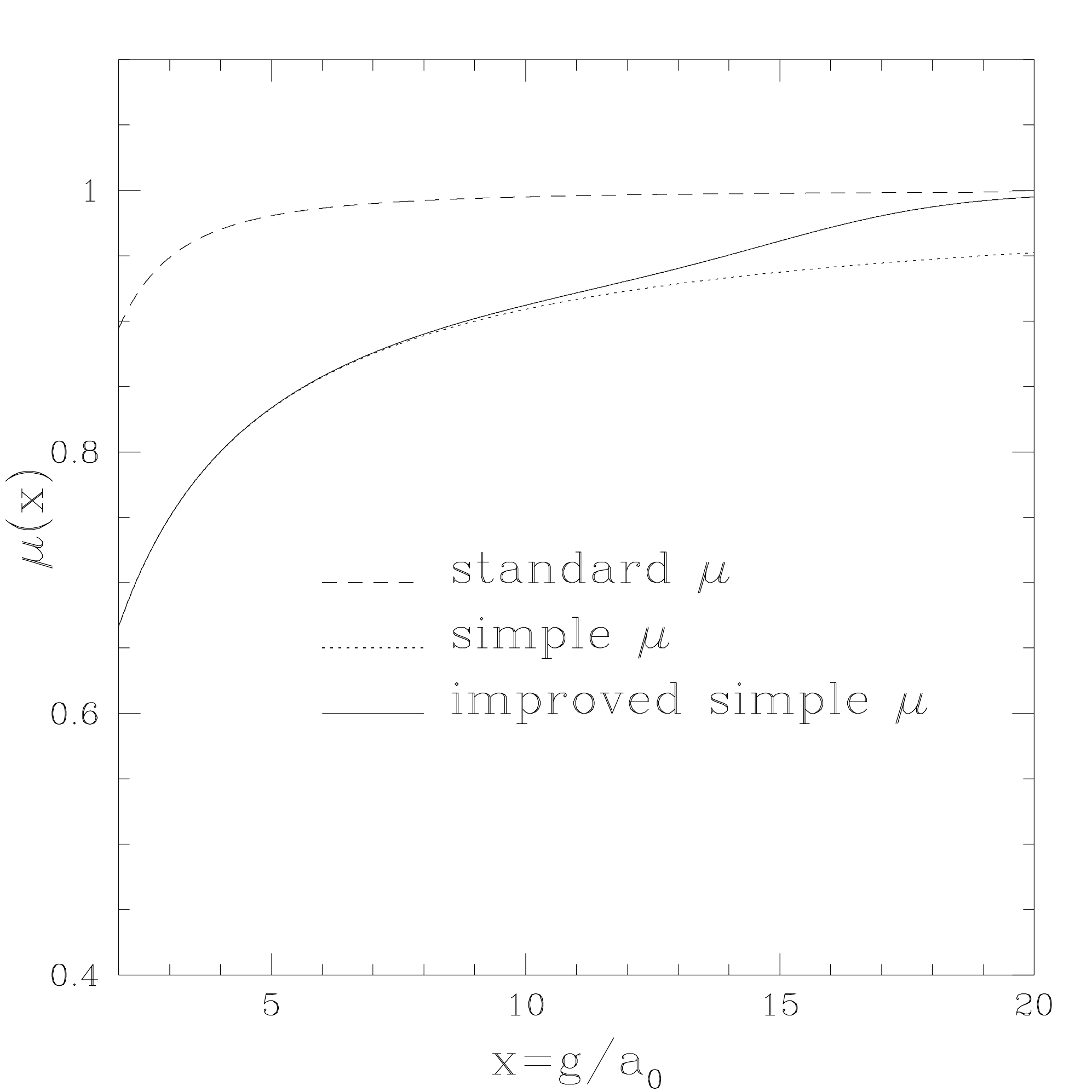}
\end{center}
\caption{The interpolating $\mu$-functions of Eq.~3 (standard, dashed line) and Eq.~4 (simple, dotted line) are displayed as a function of $x=g/a_0$. An improved simple function (solid line) interpolating between simple and standard for $x \gtrsim 10$ is also presented in order to show that a transition behaviour governed by the simple $\mu$-function in galaxies (where $x<10$) can a priori be in accordance with the Solar System constraints (where $x>>10$).
} \label{fig1}
\end{figure}

Fig.~\ref{fig1} displays those two $\mu$-functions as a function of $x$. Let us note that the simple function predicts that a constant acceleration equal to $a_0$ has to be added to the Newtonian gravitational acceleration for $g \gg a_0$. This is, for the values of $a_0$ compatible with galaxy rotation curves (Sect.~4.1), in strong disagreement with orbits of planets in the inner Solar System, and especially with measures of the perihelion precession of Mercury. A solution is to use an ``improved simple" $\mu$-function that rapidly interpolates between the simple and standard ones for
values of the gravitational acceleration $g \gtrsim 10 a_0$ (i.e. a higher value than those $g$ which are probed by galaxy rotation curves). Such an improved simple function is shown as an example on Fig.~\ref{fig1}. Nevertheless, we use the standard and simple $\mu$-functions hereafter (keeping in mind that the latter should be modified in the strong gravity regime) of Eqs.~3 and 4 in order to perform our MOND fits to THINGS galaxy rotation curves.

\begin{figure}
\begin{center}
\includegraphics[scale=0.4]{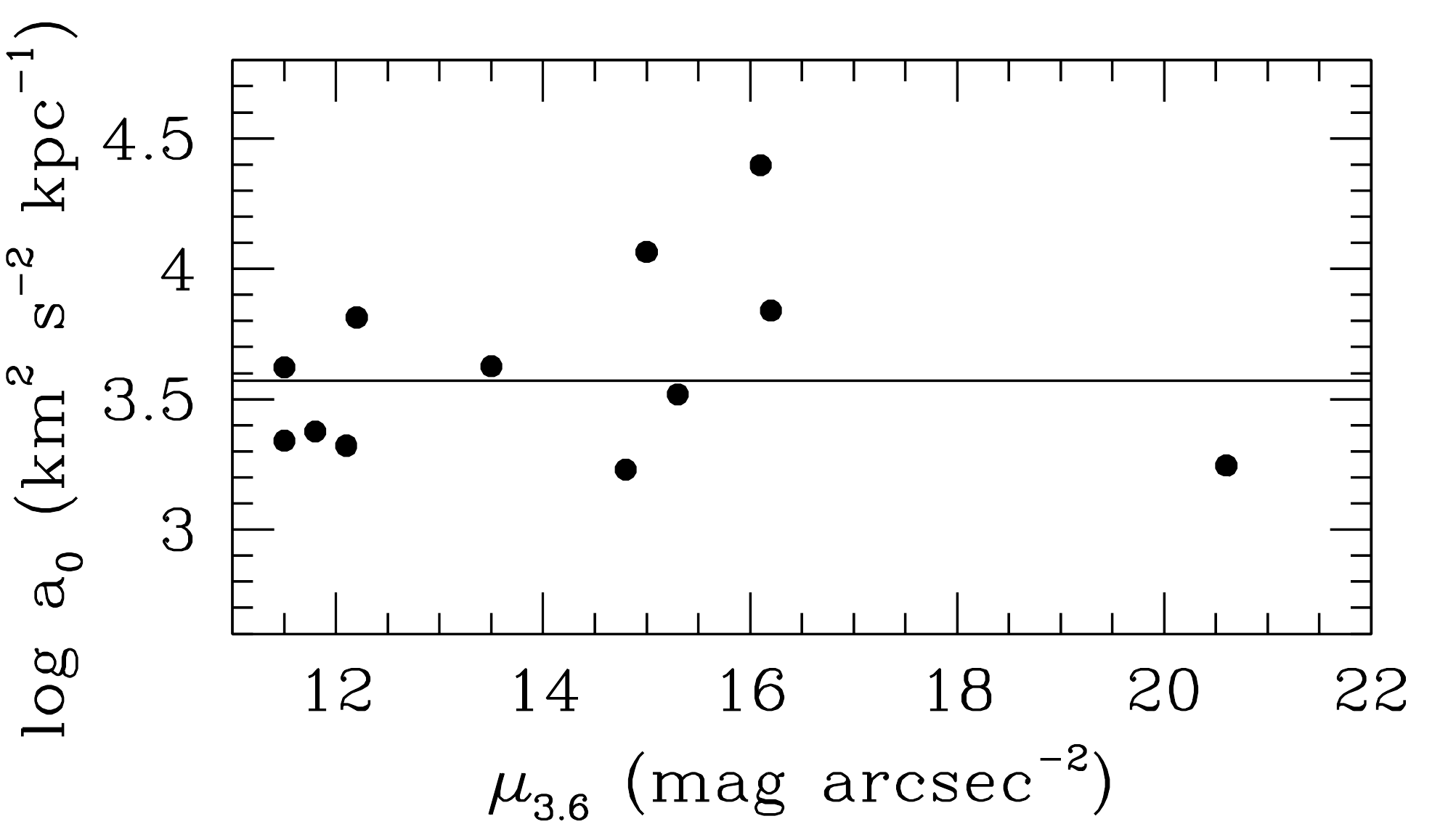}
\end{center}
\caption{Best-fit $a_0$ values (using the simple interpolating function) vs. central surface brightness in the 3.6 $\mu$m band.
} \label{a0free}
\end{figure}

\section{The sample}

We use a subset of galaxies in the THINGS survey for which rotation curves were derived in de Blok et al.~(2008). We restrict ourselves to galaxies which are not (obviously) dominated by non-circular motions.  This means we omit the bright disk galaxies NGC~3031 and NGC~4736. While not necessarily dominated by non-circular motions, we also omit NGC~2366, IC~2574 and NGC~925. These galaxies have a neutral gas distribution that 
is dominated by holes and shells, the signature of which remains visible in the radial profile of the neutral gas. In these dwarf galaxies, the neutral gas profile dominates the total radial baryonic mass distribution, 
and as the MOND prediction is derived from the observed radial surface density distribution these remaining signatures of the holes and shells could possibly lead to erroneous results. The analysis of these specific rotation curves, interesting as they might be, is left for a forthcoming paper.

This leaves us with a total of 12 galaxies: some of these have already been discussed in Bottema et al.~(2002), but with the higher resolution data available, and constrained stellar mass-to-light ratios as observed in the Spitzer IRAC $3.6 \mu {\rm m}$ band, we are able to perform a slightly more stringent test for MOND.
In fact, the 12 rotation curves that we use here are the highest quality rotation curves currently
available (in terms of spatial/spectral resolution and extent) for a sizeable sample of galaxies spanning a wide range of luminosities, and they therefore represent {an important} test for MOND or any 
theory that aims at fitting galaxy kinematics. 
 In de Blok et al. (2008) some differences between their rotation curves and those of previous publications were highlighted. These differences could be caused by the different approach taken by de Blok et al. (2008) to derive the velocity field: they fit the velocity profiles using third-order Gauss-Hermite polynomials, instead of the more conventional intensity-weighted mean. 

The two baryonic contributions to the rotation curve, necessary to compute the MOND mass model, were derived by 
de Blok et al. (2008) as follows. First, the shape (but not the amplitude) of $V_{\rm stars}$, the contribution 
of the stars to the rotation curve, was derived from the observed 3.6 $\mu$m surface brightness profile, and
slightly modified to account for the observed $(J-K)$ colour gradients as a function of radius (which are an 
indication of a radially varying stellar mass-to-light $M/L$ ratio). 
Although there might be some contamination due to young stars and hot dust, this contamination is thought to be a negligible contribution to the flux at 3.6$\mu$m, see e.g. Pahre et al. 2004, Li et al. 2007, hence the $3.6 \mu {\rm m}$ emission is considered as good a tracer of stellar mass as the more commonly used K-band (Zhu et al. 2010). 
For the vertical distribution of the stellar disk,
de Blok et al. (2008) assumed a sech$^2$ distribution with a scale height of $z_0 = h/5$, where $h$ is the radial exponential 
scale length. The amplitude of $V_{\rm stars}$ is scaled according
to the {\rm global} stellar mass-to-light $M/L$ ratio, which is left as a free parameter and then compared to the predictions
of stellar population synthesis models (e.g. Bell \& de Jong 2001). The contribution of the gaseous disk 
to the rotation curve, $V_{\rm gas}$, was derived from the observed HI surface density profiles, and then
corrected for primordial He.

The galaxy distances are determined by various methods (Cepheids, tip of the Red Giant Branch, Hubble flow, brightest stars), and their 
quoted uncertainties are often close to 10\%. They are listed in Table \ref{tab-dist}.

\begin{table}
\caption{Galaxy distances and the methods used
to determine them. References are: 1: Vink\'o et al. (2006). 2: 
Macri et al. (2001). 3: Drozdovsky \& Karachentsev (2000). 4: Karachentsev
et al. (2002). 5: Kelson et al. (1999), Freedman et al. (2001). 6: Walter et al. (2008). 7: Rawson
et al. (1997). 8: Karachentsev et al. (2004). 9: Hughes et al. (1998).}              

\label{tab-dist}      
\centering                                      
\begin{tabular}{l l l l}          
\hline\hline                        
Name     &distance(Mpc)               & Method       & Ref     \\ 
\hline                                   
NGC 2403     &$3.47\pm 0.29$               & SN, Cepheids      & 1     \\ 
NGC 2841     &$14.1\pm 1.5 $              & Cepheids      & 2     \\ 
NGC 2903     &$8.9\pm 2.2 $              & brightest stars     & 3     \\ 
NGC 2976     &$3.56\pm 0.36 $              & tip of the RGB      & 4     \\ 
NGC 3198     &$13.8\pm 1.5 $              & Cepheids      & 5     \\ 
NGC 3521     &$10.7\pm 3.2 $              & Hubble flow      & 6     \\ 
NGC 3621     &$6.64\pm 0.70 $              & Cepheids      & 7     \\ 
DDO 154      &$4.30\pm 1.07 $              & brightest stars    & 8     \\ 
NGC 5055     &$10.1\pm 3.0 $              & Hubble flow      & 6     \\ 
NGC 6946     &$5.9\pm 1.5 $              & brightest stars      & 8     \\ 
NGC 7331     &$14.72\pm1.29 $              & Cepheids      & 9     \\ 
NGC 7793     &$3.91\pm 0.39 $              & tip of the RGB      & 8     \\ 
\hline       
\end{tabular}
\tablefoot{NGC 2403 is the only galaxy whose distance differs from Walter et al. (2008). We used 
the estimate of the distance given in Vink\'o et al. (2006) because it comes from more numerous and more recent
data.}
\end{table}

\section{Results}

\subsection{The acceleration constant $a_0$}
\label{secta}

The acceleration constant $a_0$ of MOND, though unknown from first 
principles, must be the same for all galaxies, therefore it has
to be determined empirically, e.g. by fitting rotation curves.
Begeman, Broeils \& Sanders (1991) determined the value of the acceleration constant $a_0$
to be 1.21 $\times$ 10$^{-8}$ cm s$^{-2}$ from mass modelling of a number 
of nearby galaxies with the standard $\mu$-function of Eq.~3. This value was confirmed by Sanders \& Verheijen (1998)
using a sample of rotation curves of galaxies belonging to the Ursa Major 
galaxy group. However, Bottema et al. (2002) noted that using an updated value
of the distance to the Ursa Major group would bring the value of $a_0$ down
to 0.9 $\times$ 10$^{-8}$ cm s$^{-2}$.

The first fits that we performed were those with $a_0$ as a free parameter
(the stellar $M/L$ ratio being the other free parameter). The distance in these
fits was fixed at the values given in Table 1, the most accurate for each
galaxy to date, to the best of our knowledge.
We remind the reader that a fit with $a_0$ free and the distance fixed is equivalent
to a fit with the distance free and $a_0$ fixed (Begeman et al. 1991),
because the observed total gravitational acceleration is proportional to
$1/D$, where $D$ is the distance.

Swaters, Sanders \& McGaugh (2010) find a weak correlation between the R-band
central surface brightness and the best-fit value of $a_0$ as a result of making MOND mass models of 27 dwarf and low surface brightness galaxies. They find that lower surface brightness galaxies have a tendency to have lower $a_0$. In Fig. \ref{a0free} we look for a similar relation using our best-fit values of $a_0$ from the MOND fits using the simple 
$\mu$-function. We do not find the same correlation: indeed, the best-fit values of $a_0$ are scattered around 
the median value without any obvious correlation with central surface brightness. 
A thorough interpretation of the correlation (or lack
thereof) between best-fit $a_0$ and central surface brightness goes beyond the aim of this paper, but our finding might 
not invalidate the interpretation by Swaters et al. (2010) that low surface brightness galaxies could be biased
towards lower values of $a_0$ because of the external field effect (e.g. Milgrom 1983). In our sample, apart from DDO 154, 
the galaxies have a relatively high surface brightness.

The result is that the median values are (1.27 $\pm$ 0.30) $\times$ 10$^{-8}$ cm s$^{-2}$ 
for the standard $\mu$ function (eq. \ref{eqstandard}) and 
(1.22 $\pm$ 0.33) $\times$ 10$^{-8}$ cm s$^{-2}$ for the simple $\mu$ function
(eq. \ref{eqsimple}), values that are remarkably similar to the 
estimates made in previous studies (the uncertainties are calculated following
M\"uller 2000). However, for consistency we will
now use these new values in the remainder of the paper. We note that
our estimates of $a_0$ lie between $cH_0/(2 \pi) \approx 1.1 \times 10^{-8}$~cm~s$^{-2}$
(where $H_0$ is the Hubble constant) and $c \sqrt{\Lambda}/(2 \pi) \approx 1.5 \times 10^{-8}$~cm~s$^{-2}$ 
(where $\Lambda$ is the cosmological constant).
However, the estimate of $a_0$ given by Bottema et al. (2002), 0.9 $\times$ 10$^{-8}$ cm s$^{-2}$, cannot be
excluded by the present data, see Section \ref{sec_distances}.
 
\subsection{Mass-to-light ratios}

\begin{figure}
\begin{center}
\includegraphics[scale=0.4]{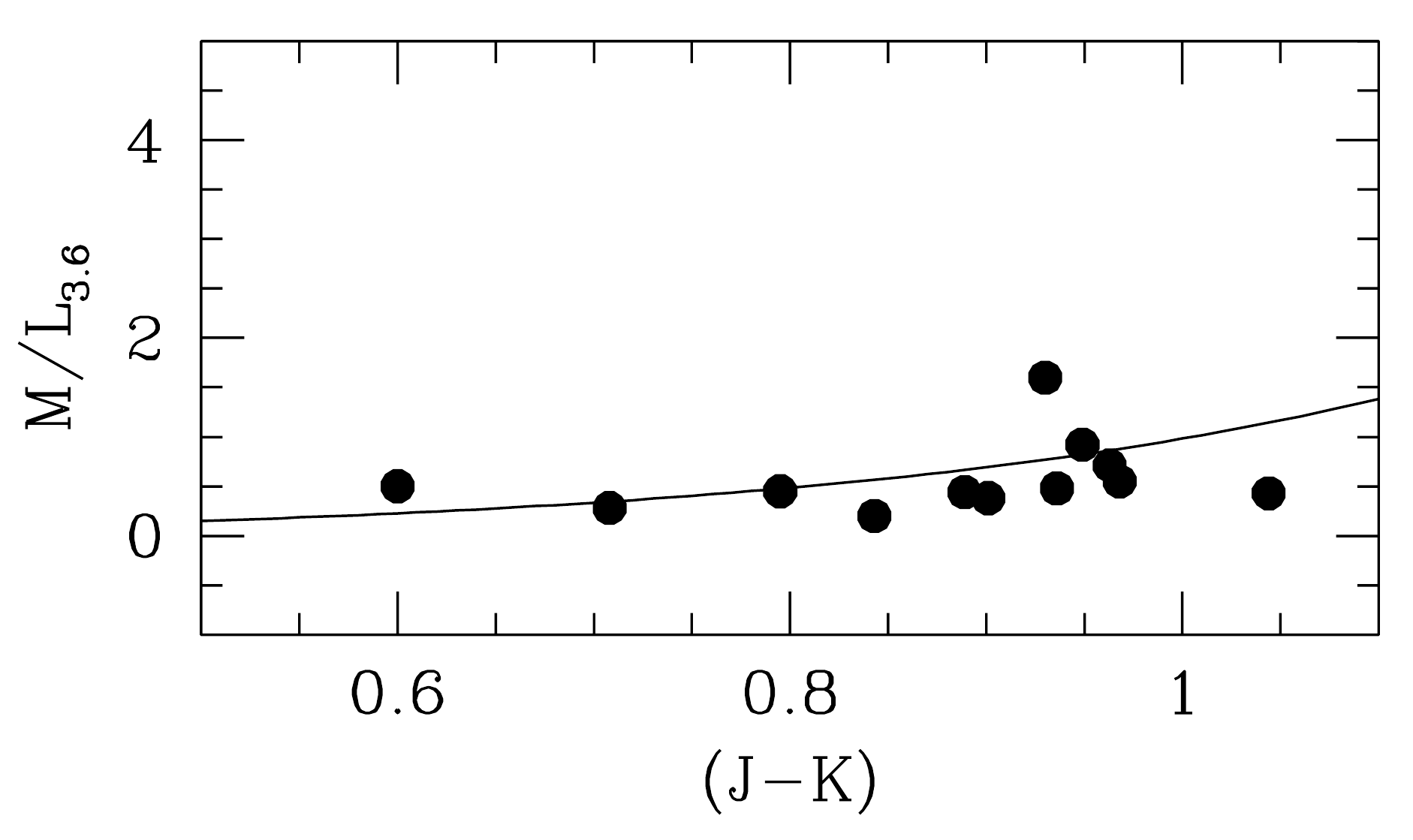}
\end{center}
\caption{Stellar $M/L$ ratio in the $3.6 \mu {\rm m}$ band vs. $(J-K)$ colour.
The full circles are the results of the MOND fits (using the simple $\mu$-function of Eq.~4 and $a_0$ = 1.22 $\times$ 10$^{-8}$ cm s$^{-2}$) with the distance
constrained within the uncertainties of its independently determined
value, whereas the solid line represents the 
predictions of stellar population synthesis models (see text for details).
} \label{colours}
\end{figure}

\begin{figure}
\begin{center}
\includegraphics[scale=0.4]{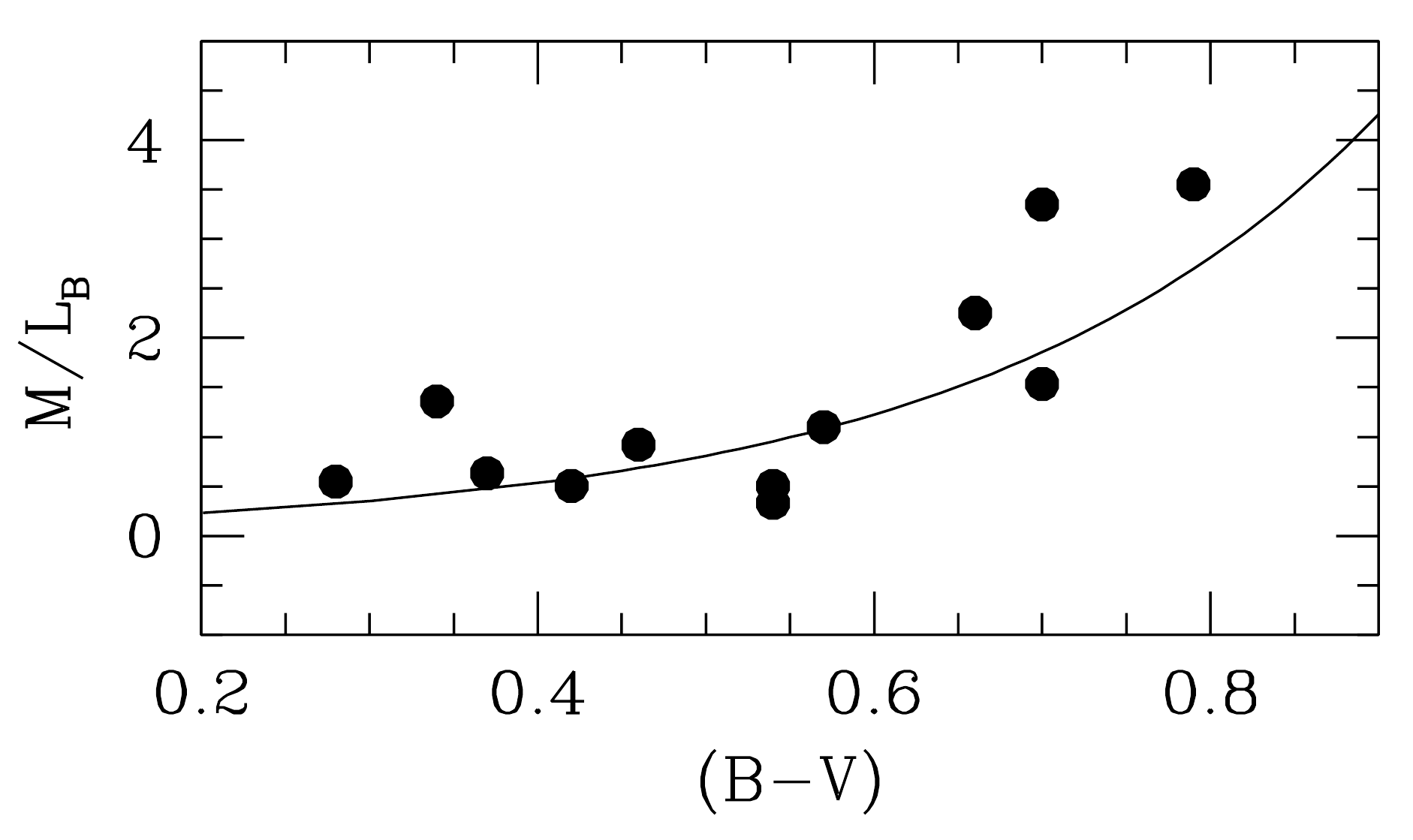}
\end{center}
\caption{Stellar $M/L$ ratio in the B band vs. $(B-V)$ colour.
See Fig. \ref{colours} for the explanation of line and symbols.
} \label{coloursb}
\end{figure}

Starting from fixed distances to the 12 galaxies, we found in Section \ref{secta} a common median value of $a_0$ corresponding to each interpolating $\mu$-function. Using this we perform 6 different types of fits to each galaxy rotation curve. For each of the two $\mu$-functions, we make fits with a fixed value of $a_0$ and (i) a fixed distance, (ii) a distance constrained to lie within the error bars from its independent determination, and (iii) a free distance with no constraints. In all cases, the stellar mass-to-light ratio of the disk (and bulge if present) is left as a free parameter. All the results are listed in Table~2. 

From the $\chi^2$ values
\footnote{\label{footnotechi}We note that in de Blok et al. (2008) the velocity difference between the approaching and receding side was considered in the error budget of each point of the rotation curve. In many cases this results in errorbars that are larger than the point-to-point scatter, which implies that the $\chi^2$ values cannot be used as probability indicators.}, 
the lack of systematic deviations and the small number of highly discrepant data points, one can 
conclude that the fits are generally good, with only a few exceptions (cf. next Section). However, one has to check that,
when a bulge is present, the mass-to-light ratio of the disk is smaller than that of the bulge, and that 
the stellar $M/L$ ratios are realistic.

There are five galaxies with a bulge. Using the standard $\mu$ function, two galaxies have the best-fit stellar
$M/L$ of the disk larger than the one of the bulge. Of the remaining three cases, one is undetermined (NGC 2903: its
best-fit $M/L$ of the bulge is zero but values larger than the best-fit $M/L$ of the disk give almost equally good fits;  
note also the likely presence of a bar, Leroy et al. 2009),
and two are realistic (the $M/L$ of the disk is smaller than $M/L$ of the bulge). On the other hand, when using the 
simple $\mu$ function no such problems arise, and the $M/L$ of the bulge is always realistic. We therefore conclude,
in line with Sanders \& Noordermeer (2007), that the simple $\mu$ function gives superior fits. For the rest of the 
paper we will thus only use the simple $\mu$ function. This justifies our proposal in Fig.~\ref{fig1} of a $\mu$ 
function that resembles the simple one at typical galactic gravitational accelerations (and the standard one for higher accelerations representative of, e.g, the Solar System).

In order to check how realistic the fitted stellar $M/L$ ratios are, we compared them with 
the results of stellar population synthesis models. In Fig. \ref{colours} we plot the best-fit ``global'' $M/L$
ratio in the $3.6 \mu {\rm m}$ band 
vs. $(J-K)$: 
\begin{equation}
M/L=\frac{(M/L)_{\rm disk} L_{\rm disk} + (M/L)_{\rm bulge} L_{\rm bulge}}{L_{\rm disk} + L_{\rm bulge}}
\end{equation}
The solid line represents the population synthesis models prediction (from Bell \& de Jong 2001, using also eq.
4 of de Blok et al. 2008) with a ``diet-Salpeter" IMF. Although with some scatter, the points lie close to the prediction,
and the $M/L$ ratios in the $3.6 \mu {\rm m}$ band vary very gradually with colour, staying constant around 0.5-1.
It is also interesting to compare our results in a band where the predicted stellar $M/L$ varies more
rapidly with colour. To achieve this, we converted our $M/L$ ratios to B-band (using the B-band luminosity
given in de Blok et al. 2008), and we made use of the 
corrected $(B-V)$ colours given in the HyperLeda database (Paturel et al. 2003). For NGC 7793, the corrected
$(B-V)$ colour was not available and we used the effective one. The results are shown in Fig. \ref{coloursb}. 
The best-fit stellar $M/L$ of MOND closely follows the predictions
of Bell \& de Jong (2001), in that redder galaxies are best fitted with a higher stellar $M/L$ ratio.

The only three galaxies where the best-fit disk $M/L$ ratio differs from the population synthesis one
by more than a factor of two are NGC 2903, NGC 2976, and NGC 7331 (see Fig. \ref{colours} and Table~2). 
In NGC 2903 the MOND fit significantly 
overpredicts the $M/L$, a phenomenon that was also observed in Newtonian mass models with dark matter
(de Blok et al. 2008). Let us note that, if one would follow population synthesis predictions, the predicted stellar disk would be surprisingly very sub-maximum
for a massive galaxy with a rapidly increasing rotation curve with maximum velocity $\sim$ 215 km s$^{-1}$.
In addition, in the central parts of NGC 2903 there is evidence for a bar (e.g. Leroy et al. 2009):
the non-circular motions associated to it further increase the uncertainties on the mass modelling
results (see also Sellwood \& Z\'anmar S\'anchez 2010).   
For NGC 2976, on the other hand, the MOND fits underpredict the stellar $M/L$ ratio; but again, this is also 
observed in the Newtonian mass models with dark matter. In addition, having a stellar disk that strongly dominates
the kinematics over most of the extent of the gaseous disk in a galaxy with maximum velocity
$\sim$ 85 km s$^{-1}$ would also be surprising. Mass model degeneracies in this galaxy 
(in particular the MOND mass models with distance free and distance constrained) are complicated by the very similar shapes of $V_{\rm stars}$ and $V_{\rm gas}$. In NGC 2976 too, Leroy et al. (2009) find an indication for a weak bar.
Also in NGC 7331 the MOND fits give a lower $M/L$ compared to the expectations from the colours; this
is the case also in the dark matter fits (see de Blok et al. 2008 where a strong dust ring is suggested as a possible explanation for the inflated stellar $M/L$ ratios predicted from the colours).

\addtocounter{table}{1} 

\subsection{Distances}
\label{sec_distances}

One then also has to check that the fits with the distances constrained to lie within the error bars from their independent determinations are of good quality. When this is not the case, it means that MOND would predict another distance than what has been measured to date. NGC~2841 is, e.g., the most well-known and most persistently problematic galaxy for MOND. Begeman et al. (1991) pointed out that a good MOND fit could only be obtained if the galaxy was a factor $\sim$ 2 further away than the Hubble distance of $\sim$ 9.5 Mpc. This large discrepancy was alleviated somewhat when HST Cepheids measurements suggested a distance of 14.1 Mpc (Macri et al. 2001), but the discrepancy remained. However, these fits were performed with the standard $\mu$-function (Eq. 3) and not the simple one (Eq. 4). Our fits here show that the problem of the distance is solved when using the simple $\mu$-function (see the reduced $\chi^2$ in Table~2), and that the stellar mass-to-light ratio is also in accordance with population synthesis models.

\begin{figure*}
\begin{center}
\includegraphics[scale=1.0]{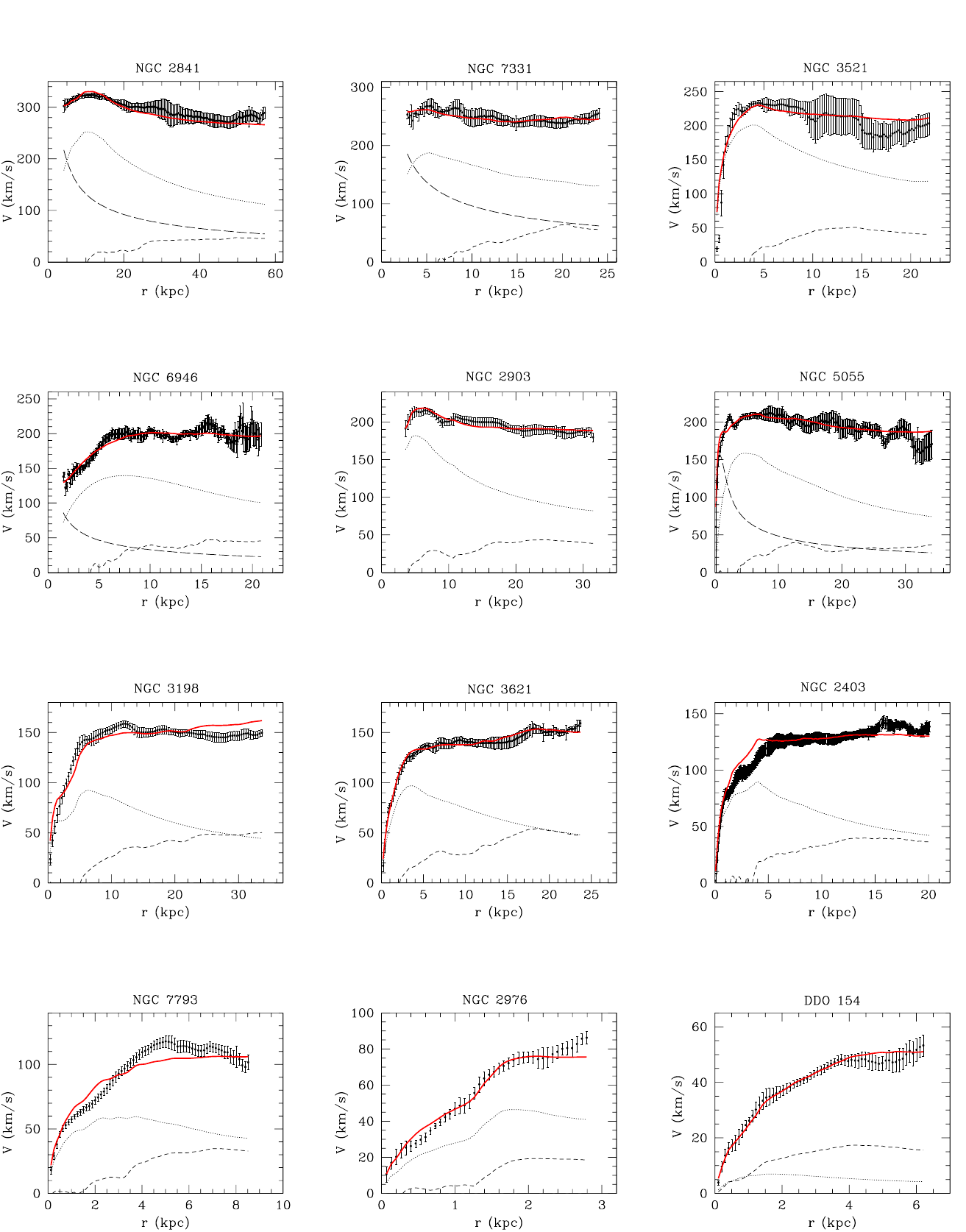}
\end{center}
\caption{Rotation curve fits with the distance ``constrained" and  $a_0=1.22 \times 10^{-8}$ ($\mu$ simple $d$ constr. in Table~2). Dashed, dotted, and long-dashed lines
represent the Newtonian contributions of the gaseous disk, stellar disk,
and bulge, respectively. The MOND best-fit model is shown as a solid red line.
} \label{fits}
\end{figure*}

All the fits of the 12 high-quality rotation curves, using the simple $\mu$-function of Eq.~4, $a_0=1.22 \times 10^{-8}$~cm~s$^{-2}$, and a distance lying within the error bars coming from an independent distance determination, are shown in Fig.~\ref{fits}. The fits are clearly very good for 9 galaxies (including NGC 3521, whose high reduced $\chi^2$ value in Table~2 is dominated by the innermost
two points, which have highly uncertain position angle and inclination, see de Blok et al. 2008).
Among the galaxies with the 3 least good fits (NGC~3198, NGC~7793, and NGC~2976), we do not discuss further 
NGC~7793 and NGC~2976, since the MOND fits present the same failures as the dark matter fits (in Newtonian
dynamics), therefore we do not consider them as evidence against MOND. We just briefly note that 
in NGC~7793 the value of the inclination angle fitted by de Blok et al. (2008) is low and 
presents large variations in adjacent radii, which results in a poorly constrained rotation curve;  
in NGC~2976 the amplitude of the non-circular motions (Trachternach et al. 2008) 
is correlated with the amplitude of the fit residuals.

\begin{figure*}
\begin{center}
\includegraphics[scale=1.0]{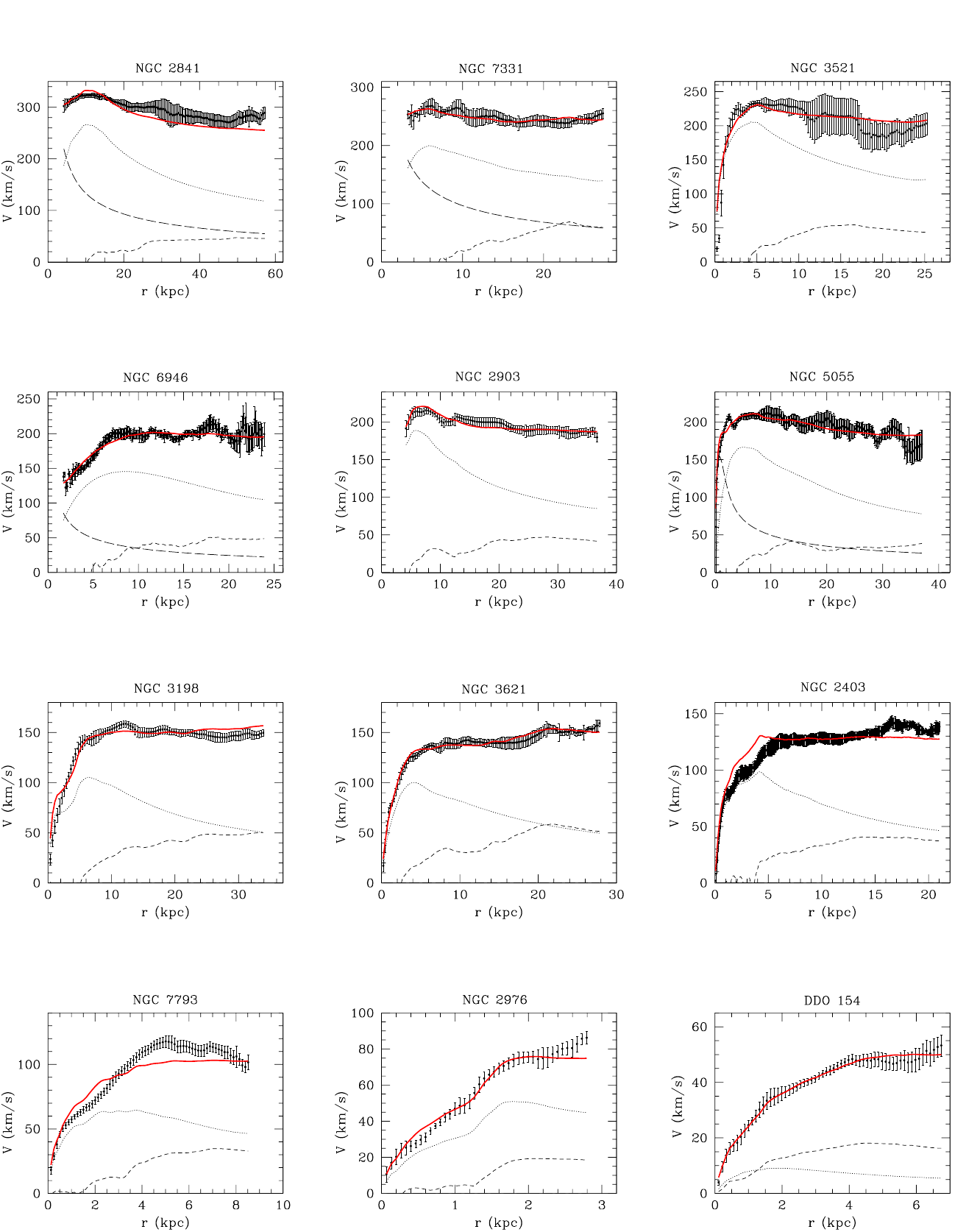}
\end{center}
\caption{Rotation curve fits with the distance constrained to lie within the error bars of Table \ref{tab-dist}. and  $a_0=0.9 \times 10^{-8}$ cm s$^{-2}$. 
The lines are described in Fig. \ref{fits}.
} \label{fits_0p9}
\end{figure*}

Before proceeding with a detailed analysis of possible problems with the rotation curve of NGC 3198 in Sect. \ref{sec_3198}, we finally consider the possibility that the true value of $a_0$ for all galaxies is actually at the lower end of our best-fit interval of Sect. \ref{secta}, i.e. a value compatible with the one determined by Bottema et al. (2002). As a matter of fact, a good reason for this is that the Ursa Major (UMa) galaxy group (e.g., Sanders \& Verheijen 1998, Gentile, Zhao \& Famaey 2008) is nowadays thought to be at a distance of 18.6 Mpc (Tully \& Pierce 2000), implying a best-fit value of $a_0$ close to the one of Bottema et al. (2002), see e.g. Gentile et al. (2008). To get as good fits to the rotation curves of UMa galaxies as those obtained with $a_0=0.9 \times 10^{-8}$ cm s$^{-2}$ with a higher value of the order of $a_0=1.2 \times 10^{-8}$ cm s$^{-2}$, the distance of the group should be smaller, of the order of 15 Mpc (as originally assumed by Sanders \& Verheijen 1998). For this reason, we plot in Fig. \ref{fits_0p9} the fits of the 12 rotation curves using $a_0=0.9 \times 10^{-8}$ cm s$^{-2}$ (and still the simple $\mu$-function and the distance constrained to lie within the error bars of Table \ref{tab-dist}). As can be seen, the fits remain of approximately the same quality, apart for 3 galaxies: NGC 2841 and NGC 2403 have worse fits \footnote{In this case the less well fitted galaxy would be NGC 2403. Possible effects could be the fact that the $r$,  J, H and K photometric profiles have a different shape from the $3.6 \mu {\rm m}$ band (see Kent 1987, Fraternali et al. 2002, de Blok et al. 2008). In addition, Fig. 7 of de Blok et al. (2008) shows that the outer parts of the rotation curve are quite uncertain. In that paper, the stellar component is also modelled with two separate components: the use of two different disks for the stellar contribution does not change significanly the results., but the quality of the fit of NGC 3198 improves.} 

\begin{figure}
\begin{center}
\includegraphics[scale=0.4]{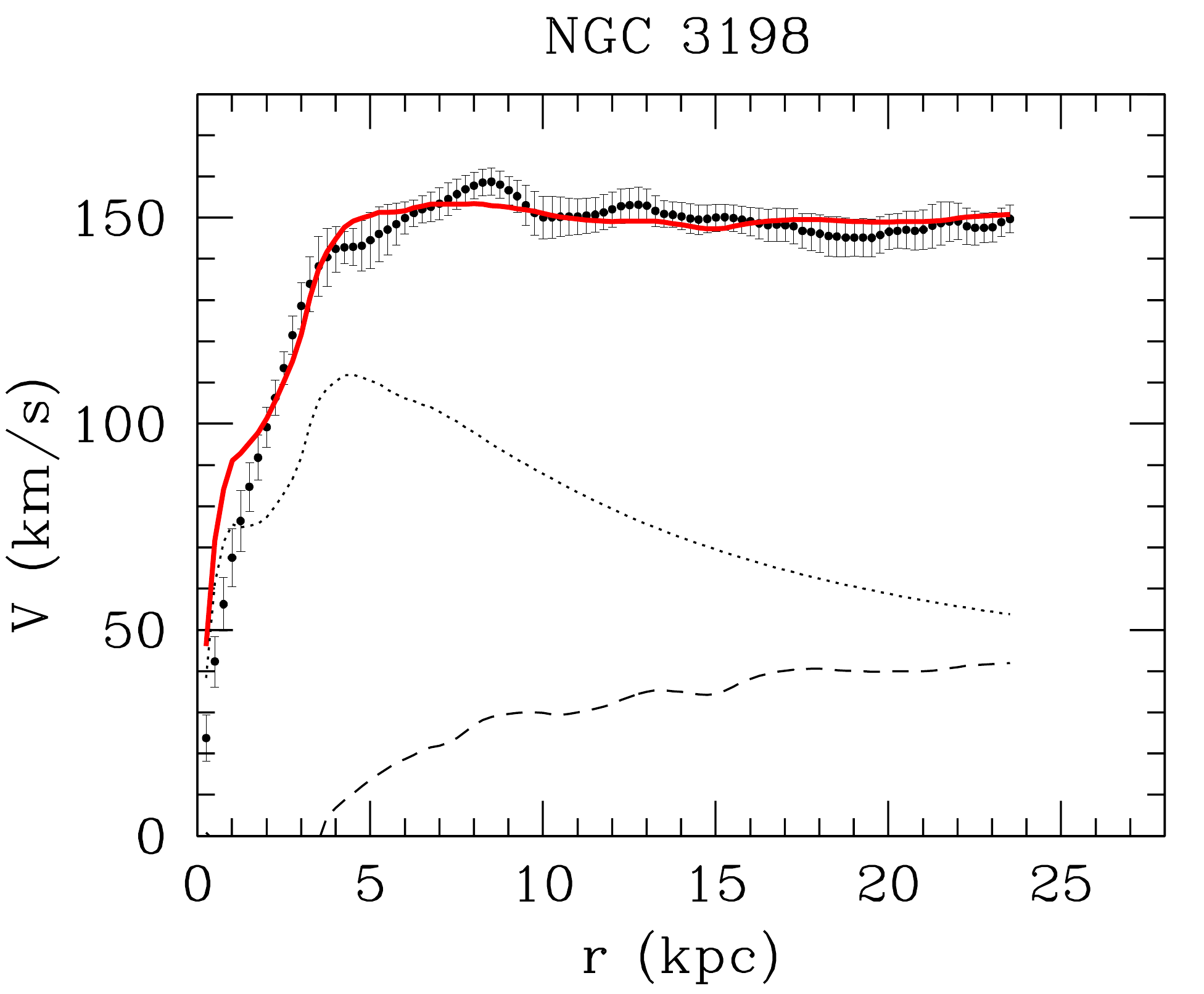}
\end{center}
\caption{Rotation curve fit of NGC 3198 with $a_0=1.2 \times 10^{-8}$ cm s$^{-2}$ the distance as a free parameter ($\mu$ simple $d$ free in Table~2).
The distance is 8.6 Mpc and the stellar mass-to-light ratio in the 3.6$\mu$m band is 1.01.
The lines are described in Fig. \ref{fits}.
} \label{3198}
\end{figure}

\begin{figure}
\begin{center}
\includegraphics[scale=0.4]{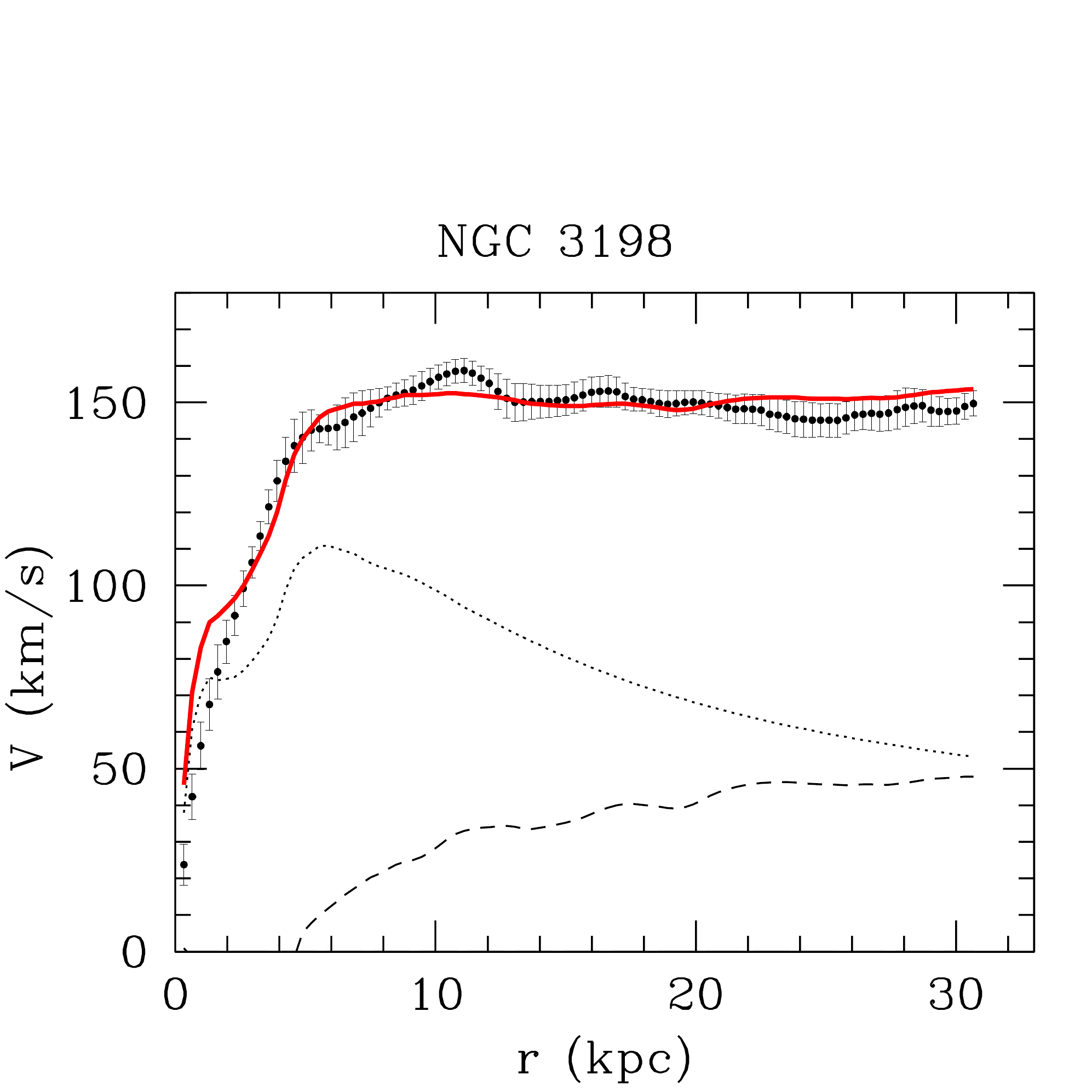}
\end{center}
\caption{Rotation curve fit of NGC 3198 with $a_0=0.9 \times 10^{-8}$ cm s$^{-2}$, a distance of 11.2 Mpc (see Section \ref{sec_3198}), a best-fit stellar M/L ratio in the 3.6$\mu$m band  of 0.76, and the simple $\mu$ function.
The lines are described in Fig. \ref{fits}.
} \label{3198_11p2}
\end{figure}

\subsection{NGC 3198} 
\label{sec_3198}

An excellent fit can be found with $a_0=1.2 \times 10^{-8}$ cm s$^{-2}$ by leaving the distance unconstrained. As already
noted in previous studies (e.g. Bottema et al. 2002), MOND prefers a smaller distance. Fig.~\ref{3198}
shows a very good fit with a distance of 8.6 Mpc, which is significantly lower than the Cepheids-based
one of 13.8 Mpc. The distance that one would get by
fixing the stellar $M/L$ ratio to the population synthesis value is also lower than the Cepheids value (9.6 Mpc).
We note that other methods (the Tully-Fisher distance or the Hubble flow distance, both less accurate than 
the Cepheids) also yield lower values (see van Albada et al. 1985 and Bottema et al. 2002) than the Cepheids.
It is however also interesting to note that the regions where the fit with the 
distance constrained is most discrepant (roughly, the first and last thirds of the rotation curve)
are also those where the amplitude of the non-circular motions is higher, 
taken from Trachternach et al. (2008) and they are about of the same order of magnitude.
The increase of the non-circular motions from the middle part of the rotation curve to the 
outer parts at $\sim10$ km s$^{-1}$ is also noted in Sellwood \& Z\'anmar S\'anchez (2010). 
The interpretation is further complicated by the fact that the IRAC 3.6 $\mu$m image shows what seems to be
an end-on bar in the very inner part of NGC 3198. The main spiral arms of NGC 3198 originate here, and
it is not clear how these affect the dynamics. The use of two different disks for the stellar contribution (see de Blok et al. 2008) does not change significanly
the results.
We thus conclude that a full MOND modelling of the two-dimensional velocity field of this galaxy, taking into account the full modified Poisson equation(s) of Bekenstein \& Milgrom~(1984) or Milgrom~(2010), as well as the non-axisymmetry of the galactic potential, would lead to a benchmark test for the viability of MOND as a modification of gravity. 
We note, however, that a lower value of $a_0$would improve the fit of NGC 3198 although the fits of NGC 2841 and NGC 2403 would get worse. With $a_0=0.9 \times 10^{-8}$ cm s$^{-2}$, a fit equivalent to the one of Fig. ~\ref{3198} would be obtained for a distance of 10.3 Mpc instead of 8.6 Mpc, closer to the Cepheid-based distance of 13.8 Mpc. Actually, it has been mentioned in the literature that there might be a reddening problem in determining the distance. In Macri et al. (2001), the H-band Cepheids distance of NGC 3198 is 11.2 Mpc. So, with $a_0=0.9 \times 10^{-8}$ cm s$^{-2}$, there might be no problem with the rotation curve of NGC 3198 (see Fig.~\ref{3198_11p2}).

\subsection{Scaling relations}

An interesting way to interpret the above MOND results is to phrase them in terms of the usual dark matter framework, considering MOND as a phenomenological, empirical, law encompassing the behavior of dark matter in galaxies. The additional gravity generated by MOND, compared to the Newtonian case, can indeed be attributed to what one would call dark matter in the Newtonian context, and this effective matter is called ``phantom dark matter". For the gravity generated by baryons, we hereafter use the mass-to-light ratios from the fits made with the simple interpolation function and the distance constrained.

\begin{figure}
\begin{center}
\includegraphics[scale=0.45]{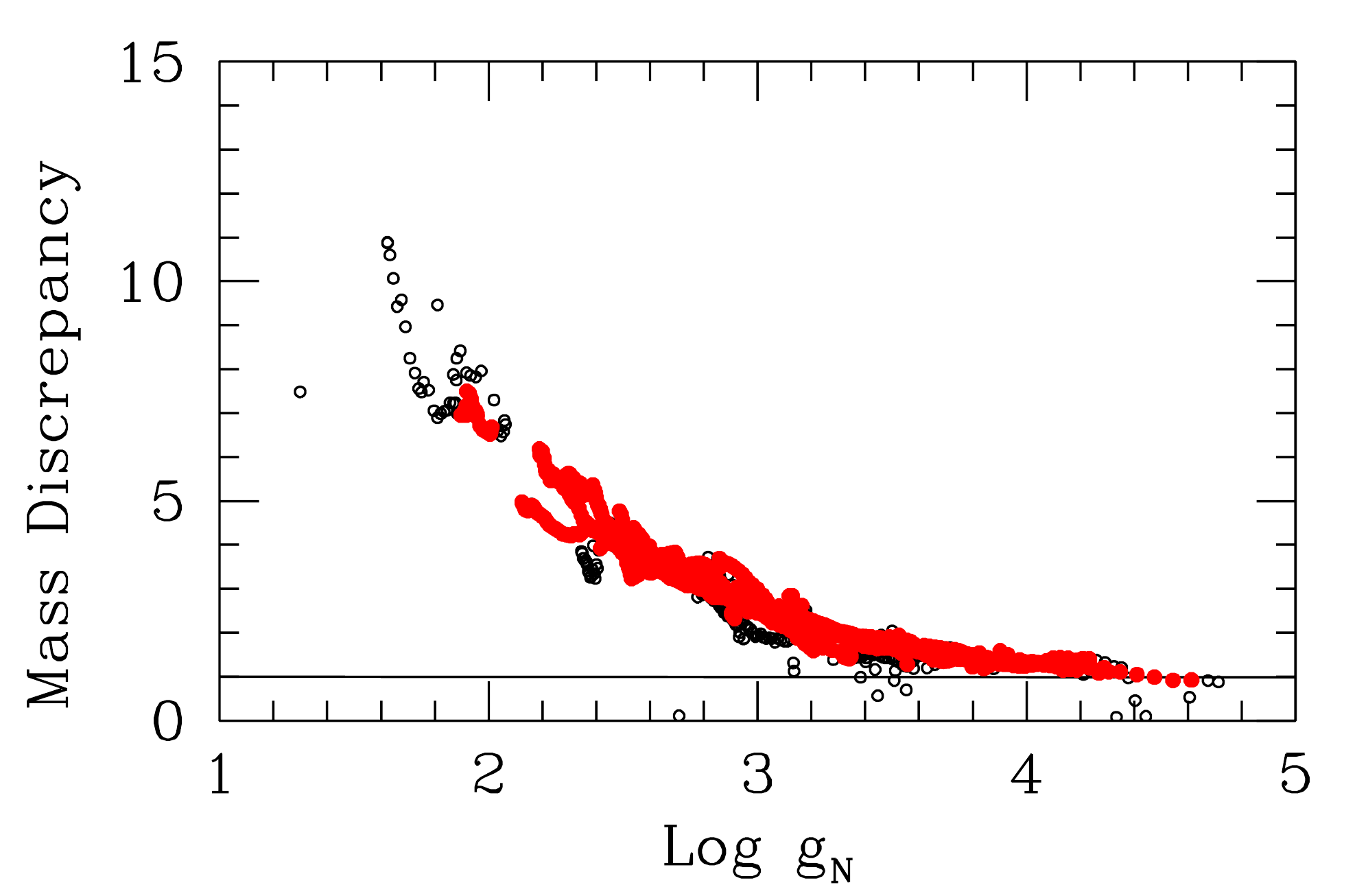}
\end{center}
\caption{Mass discrepancy-acceleration relation using the rotation curve data of our sample. 
The gravitational acceleration generated by baryons ($g_{\rm N}$) is
measured in km$^{2}$ s$^{-2}$ kpc$^{-1}$ and they result from the fits
made with the simple interpolation function and the distance constrained in Table 2.
Black (open) circles represent the data points with an uncertainty larger than
5\%. The data points with an uncertainty smaller than 5\% are shown
as red (full) circles.
} \label{mda_things_gn}
\end{figure}

In Fig.~\ref{mda_things_gn}, we display the scaling relation known as the Mass Discrepancy-Acceleration relation (McGaugh~2004), showing that the ratio of enclosed total dynamical mass (in Newtonian gravity) w.r.t. enclosed baryonic mass at any radius is a function of the gravity generated by the baryons at this radius. This relation precisely traces the $\mu$-function of MOND, and the small scatter around the line indicate the deviation from the MOND behavior, mostly consistent with observational errors.

\begin{figure}
\begin{center}
\includegraphics[scale=0.45]{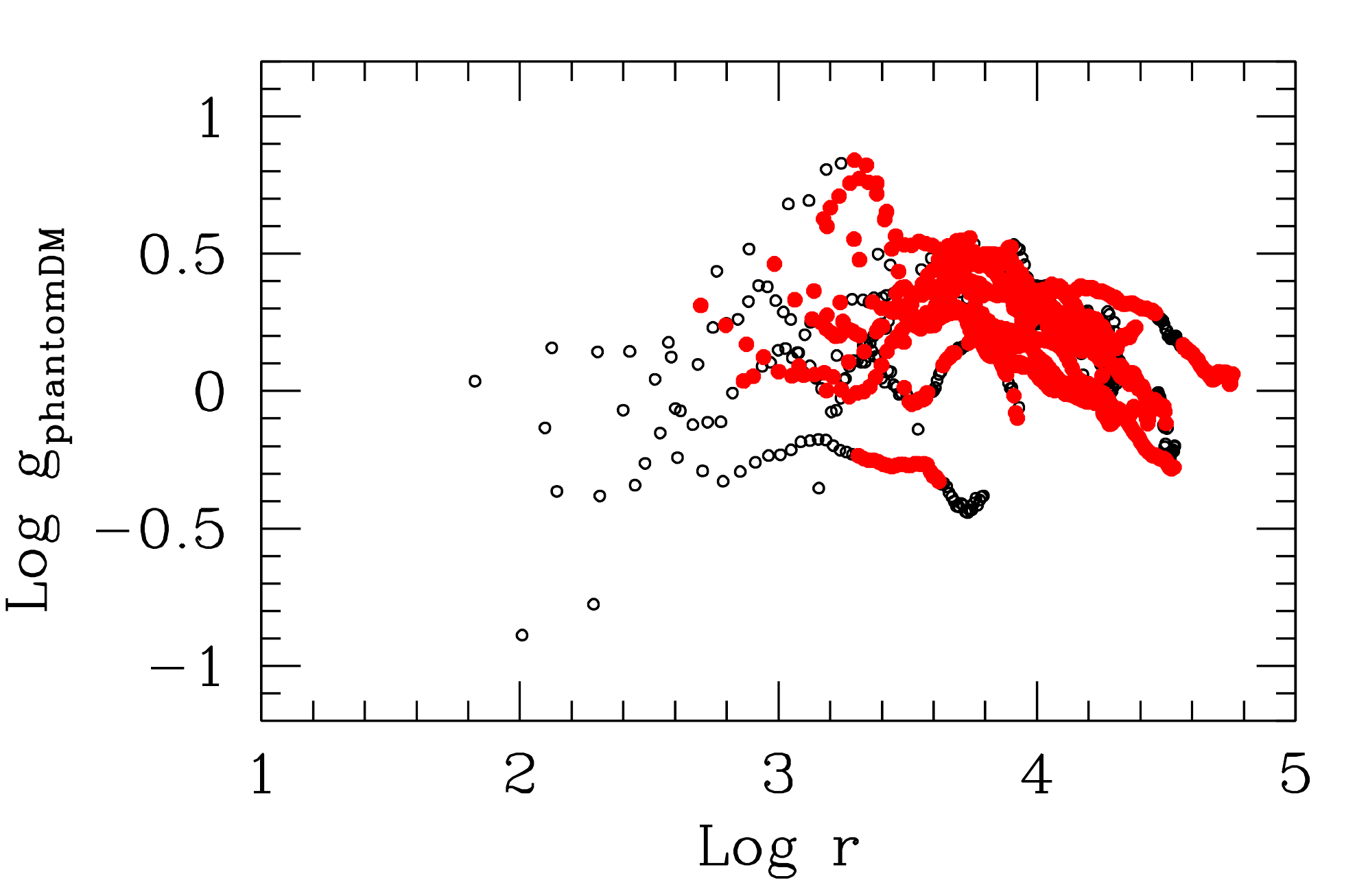}
\end{center}
\caption{Gravitational acceleration generated by phantom dark matter ($g_{\rm phantomDM}$,
measured in km$^{2}$ s$^{-2}$ pc$^{-1}$) versus radius (in pc). The values of $g_{\rm phantomDM}$
result from the fits made with the simple interpolation function and the distance constrained in Table 2.
} \label{acc_vs_r}
\end{figure}

\begin{figure}
\begin{center}
\includegraphics[scale=0.45]{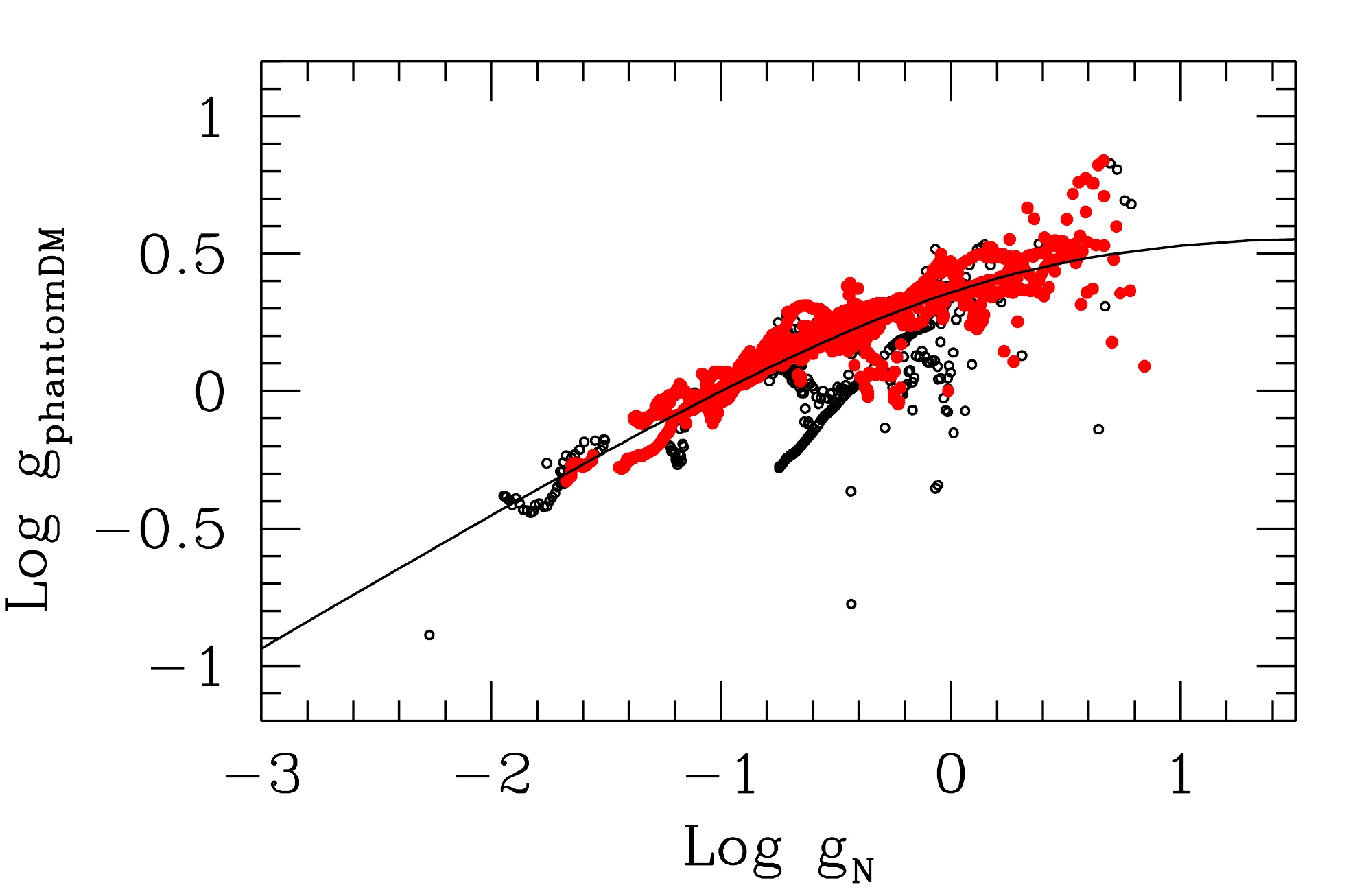}
\end{center}
\caption{Gravitational acceleration generated by phantom dark matter ($g_{\rm phantomDM}$)
versus gravitational acceleration generated by baryons ($g_{\rm N}$). The units are 
km$^{2}$ s$^{-2}$ pc$^{-1}$. The values of $g_{\rm phantomDM}$ and $g_{\rm N}$
result from the fits made with the simple interpolation function and the distance constrained in Table 2.
The solid line represents the simple interpolation function of MOND (eq. \ref{eqsimple})
} \label{phantom_gravity_gn}
\end{figure}

Then, following Walker et al.~(2010), we plot the gravity of (phantom) dark matter as a function of radius for the 12 galaxies of the sample. This is plotted on Fig.~\ref{acc_vs_r}. We remarkably find that the additional gravity predicted by MOND is in accordance with the mean and scatter in Fig.~1 of Walker et al.~(2010). However, if one plots (Fig.~\ref{phantom_gravity_gn}) this additional gravity as a function of  the baryonic {\it gravitational acceleration}, the scatter is much lower and samples the $\mu$-function. Since for the considered range of such gravitational accelerations, the range of ``phantom dark matter" gravities is not very large, it gives the illusion of a dark matter gravity which is more or less constant with radius.

\section{Conclusion}

We re-analysed the ability of the Modified Newtonian Dynamics (MOND; Milgrom 1983) paradigm to fit galaxy rotation curves, using the most up-to-date high-resolution HI data for nearby ($d<15$~Mpc) galaxies from the recent THINGS survey (Walter et al. 2008). We selected a subset of 12 galaxies not obviously dominated by non-circular motions, and yielding the most reliable mass models.

First, we redetermined the value of the acceleration parameter in MOND ($a_0$), 
which is unknown a priori but has to be the same for all galaxies.
This was done for both commonly used interpolating functions $\mu$ of MOND.
We find a median value of  $a_0=(1.27$ $\pm$ 0.30) $\times$ 10$^{-8}$ cm s$^{-2}$ for
the ``standard" $\mu$ function (Eq.~3), and $a_0=(1.22$ $\pm$ 0.33) $\times$ 10$^{-8}$ cm s$^{-2}$ for the ``simple" $\mu$ function (Eq.~4), very
close to the value that had been determined in previous studies (e.g.
Begeman et al. 1991). 

Then, fixing these values for $a_0$, we performed three fits for each
$\mu$ function: with the distance fixed at the value determined in an
independent way, then by leaving the distance free but constrained within the uncertainties
of this distance determination, and then with the distance as a free
parameter with no constraints (Table~2). We find that the MOND fits with
the distance ``constrained" are of very good quality (Fig.~\ref{fits}), with three exceptions: two of these are galaxies that
cannot give good fits using Newtonian dynamics plus dark matter (NGC 2976 and NGC 7793) either, see de Blok et al. (2008).  
For the remaining galaxy (NGC 3198) there is indeed some tension between observations and the MOND fit, 
that might be explained by the presence of non-circular motions, a small distance (see Fig. \ref{3198}), or a value of $a_0$ at the lower end of our best-fit interval (see Fig. \ref{3198_11p2}). In any case, further observations (constraining the distance) and modelling of NGC 3198 in the MOND context should thus lead to a benchmark test for MOND as a modification of gravity. But we also show that MOND, as an empirical law encompassing the behavior of the gravitational field on galaxy scales, whatever its cause, is still very successful and summarizes old and new scaling relations with a remarkable consistency (Sect.~4.5).

We also conclude that, both from arguments of best-fit stellar mass-to-light ratios (Sect~4.2) and best-fit distances (Sect.~4.3), the simple $\mu$-function is preferred over the standard one. As noted by Famaey \& Binney~(2005) and McGaugh~(2008), this is also the case when fitting the terminal velocity curve of our own Milky Way galaxy. Angus, Famaey \& Diaferio (2010) also reached the same conclusion from using temperature profiles of 
the X-ray emitting gas of a sample of clusters, and 
from assuming that dark matter in MONDian galaxy 
clusters is made of 11eV fermionic particles\footnote{Let us note that such a hot dark matter component could also play a role in the strong and weak gravitational lensing of elliptical galaxies (Ferreras et al. 2009, Tian et al. 2009)}.

Let us however note that,  in order to constrain $\mu$ from galaxy rotation curves within the modified gravity framework of MOND (see Sect.~2), one should actually calculate predictions of the modified Poisson formulations of Bekenstein \& Milgrom~(1984) or Milgrom~(2010) numerically for each galaxy model, and for each choice of parameters. Our present conclusion for THINGS galaxy rotation curves does hold only for the modified inertia formulation for circular orbits given here by Eq.~1.

\section*{Acknowledgements}

GG is a postdoctoral researcher of the FWO-Vlaanderen (Belgium). BF is a CNRS Senior Resarch Associate (France) and 
acknowledges the support of the AvH foundation (Germany).
The work of WJGdB is based upon research supported by the South African Research Chairs Initiative 
of the Department of Science and Technology and National Research Foundation.
We acknowledge the usage of the HyperLeda database (http://leda.univ-lyon1.fr).
We are greatful to the whole THINGS (The HI Nearby Galaxy Survey) team, and in particular to 
Fabian Walter and Elias Brinks for valuable suggestions. We also
wish to thank Garry Angus for very useful comments on the paper.
Finally, we thank the referee, Bob Sanders,
for valuable comments that improved the quality and scientific content of this paper.

\longtab{2}{
\begin{longtable}{llccccccc}    
\caption{Mass modelling results. $M/L_{\rm disk}$ and $M/L_{\rm bulge}$ are in the $3.6 \mu {\rm m}$ band.
The values called $f$ are the ratio between the best-fit value and the ``nominal'' one (stellar $M/L$ from
the colours and the independently determined distance). The distance is in Mpc and $\chi^2_{\rm red}$ is the reduced $\chi^2$ 
(this number is useful to compare different models of the same galaxy but should not be used as a true probability indicator, see footnote \ref{footnotechi}). The values refer to the fits made with $a_0$ = 1.22 $\times$ 10$^{-8}$ cm s$^{-2}$.}\\
\hline\hline
Name & fit & $M/L_{\rm disk}$ & $f_{\rm disk}$ &
$M/L_{\rm bulge}$ & $f_{\rm bulge}$ & dist & $f_{\rm dist}$ &
$\chi^2_{\rm red}$ \\
\hline
\endfirsthead
\caption{continued.}\\
\hline\hline
Name & fit & $M/L_{\rm disk}$ & $f_{\rm disk}$ &
$M/L_{\rm bulge}$ & $f_{\rm bulge}$ & dist & $f_{\rm dist}$ &
$\chi^2_{\rm red}$ \\
\hline
\endhead
\hline
\endfoot

NGC 2841 &$\mu$ stand. $d$ fixed   & $1.46$ & $1.97$ & $0.87$& $1.04$& 14.10                  & -                       & 2.23     \\ 
         &$\mu$ stand. $d$ constr. & $1.26$ & $1.70$ & $0.89$& $1.06$&$15.60$ & $1.11$  & 1.41     \\ 
         &$\mu$ stand. $d$ free    & $0.81$ & $1.09$ & $0.88$& $1.05$&$20.74$ & $1.47$  & 0.42     \\  
         &$\mu$ simple $d$ fixed   & $1.04$ & $1.41$ & $1.10$& $1.31$& 14.10                  & -                       & 1.32     \\ 
         &$\mu$ simple $d$ constr. & $0.89$ & $1.20$ & $1.04$& $1.24$&$15.60$ & $1.11$  & 0.87     \\ \vspace{0.048cm}
         &$\mu$ simple $d$ free    & $0.52$ & $0.70$ & $0.86$& $1.02$&$21.55$ & $1.53$  & 0.23     \\  
NGC 7331 &$\mu$ stand. $d$ fixed   & $0.52$ & $0.74$ & $0.94$& $0.94$& 14.72                  & -                       & 0.34     \\ 
         &$\mu$ stand. $d$ constr. & $0.62$ & $0.89$ & $0.81$& $0.81$&$13.43$ & $0.91$  & 0.26     \\  
         &$\mu$ stand. $d$ free    & $0.68$ & $0.97$ & $0.72$& $0.72$&$12.78$ & $0.87$  & 0.25     \\ 
         &$\mu$ simple $d$ fixed   & $0.33$ & $0.47$ & $1.24$& $1.24$& 14.72                  & -                       & 0.43     \\ 
         &$\mu$ simple $d$ constr. & $0.40$ & $0.57$ & $1.22$& $1.22$&$13.43$ & $0.91$  & 0.34     \\ \vspace{0.048cm}
         &$\mu$ simple $d$ free    & $0.64$ & $0.91$ & $1.16$& $1.16$&$10.39$ & $0.71$  & 0.23     \\ 
NGC 3521 &$\mu$ stand. $d$ fixed   & $0.58$ & $0.79$ & -                     & -                     & 10.70                  & -                       & 6.32     \\ 
         &$\mu$ stand. $d$ constr. & $0.75$ & $1.03$ & -                     & -                     & $8.68$ & $0.81$  & 6.19     \\  
         &$\mu$ stand. $d$ free    & $0.75$ & $1.03$ & -                     & -                     & $8.68$ & $0.81$  & 6.19     \\ 
         &$\mu$ simple $d$ fixed   & $0.44$ & $0.60$ & -                     & -                     & 10.70                  & -                       & 5.84     \\ 
         &$\mu$ simple $d$ constr. & $0.71$ & $0.97$ & -                     & -                     & $7.50$ & $0.70$  & 5.49     \\ \vspace{0.048cm}
         &$\mu$ simple $d$ free    & $0.79$ & $1.08$ & -                     & -                     & $6.91$ & $0.65$  & 5.48     \\  
NGC 6946 &$\mu$ stand. $d$ fixed   & $0.60$ & $0.94$ & $0.60$& $0.60$&  5.90                  & -                       & 1.04     \\ 
         &$\mu$ stand. $d$ constr. & $0.50$ & $0.78$ & $0.61$& $0.61$& $6.60$ & $1.12$  & 1.00     \\  
         &$\mu$ stand. $d$ free    & $0.50$ & $0.78$ & $0.61$& $0.61$& $6.60$ & $1.12$  & 1.00     \\  
         &$\mu$ simple $d$ fixed   & $0.42$ & $0.66$ & $0.61$& $0.61$&  5.90                  & -                       & 1.02     \\ 
         &$\mu$ simple $d$ constr. & $0.37$ & $0.58$ & $0.55$& $0.55$& $6.41$ & $1.09$  & 1.00     \\  \vspace{0.048cm}
         &$\mu$ simple $d$ free    & $0.37$ & $0.58$ & $0.55$& $0.55$& $6.41$ & $1.09$  & 1.00     \\
NGC 2903
\footnote{In NGC 2903 the best-fit $M/L$ of the bulge is zero, but values larger than the best-fit $M/L$ of the disk give almost equally good fits.}    &$\mu$ stand. $d$ fixed   & $2.57$ & $4.21$ & $0.00$& $0.00$&  8.90                  & -                       & 1.03     \\ 
         &$\mu$ stand. $d$ constr. & $2.30$ & $3.77$ & $0.00$& $0.00$& $9.55$ & $1.07$  & 0.94     \\ 
         &$\mu$ stand. $d$ free    & $2.30$ & $3.77$ & $0.00$& $0.00$& $9.55$ & $1.07$  & 0.94     \\  
         &$\mu$ simple $d$ fixed   & $1.92$ & $3.15$ & $0.00$& $0.00$&  8.90                  & -                       & 0.63     \\ 
         &$\mu$ simple $d$ constr. & $1.71$ & $2.80$ & $0.00$& $0.00$& $9.56$ & $1.07$  & 0.58     \\ \vspace{0.048cm}
         &$\mu$ simple $d$ free    & $1.71$ & $2.80$ & $0.00$& $0.00$& $9.56$ & $1.07$  & 0.58     \\ 
NGC 5055 &$\mu$ stand. $d$ fixed   & $0.43$ & $0.54$ & $0.46$& $0.35$& 10.10                  & -                       & 2.63     \\ 
         &$\mu$ stand. $d$ constr. & $0.75$ & $0.95$ & $0.55$& $0.42$& $7.07$ & $0.70$  & 0.97     \\ 
         &$\mu$ stand. $d$ free    & $0.84$ & $1.06$ & $0.57$& $0.44$& $6.55$ & $0.65$  & 0.91     \\  
         &$\mu$ simple $d$ fixed   & $0.30$ & $0.38$ & $0.43$& $0.33$& 10.10                  & -                       & 1.80     \\ 
         &$\mu$ simple $d$ constr. & $0.55$ & $0.70$ & $0.56$& $0.43$& $7.07$ & $0.70$  & 0.86     \\ \vspace{0.048cm}
         &$\mu$ simple $d$ free    & $0.66$ & $0.84$ & $0.61$& $0.47$& $6.27$ & $0.62$  & 0.80     \\ 
NGC 3198 &$\mu$ stand. $d$ fixed   & $0.49$ & $0.61$ & -                     & -                     & 13.80                  & -                       & 6.27     \\ 
         &$\mu$ stand. $d$ constr. & $0.67$ & $0.84$ & -                     & -                     &$12.30$ & $0.89$  & 3.81     \\ 
         &$\mu$ stand. $d$ free    & $1.43$ & $1.79$ & -                     & -                     & $8.38$ & $0.61$  & 1.56     \\ 
         &$\mu$ simple $d$ fixed   & $0.37$ & $0.46$ & -                     & -                     & 13.80                  & -                       & 6.18     \\ 
         &$\mu$ simple $d$ constr. & $0.48$ & $0.60$ & -                     & -                     &$12.30$ & $0.89$  & 3.79     \\ \vspace{0.048cm}
         &$\mu$ simple $d$ free    & $1.01$ & $1.26$ & -                     & -                     & $8.60$ & $0.62$  & 1.32     \\  
NGC 3621 &$\mu$ stand. $d$ fixed   & $0.51$ & $0.86$ & -                     & -                     &  6.64                  & -                       & 0.70     \\ 
         &$\mu$ stand. $d$ constr. & $0.60$ & $1.02$ & -                     & -                     & $6.14$ & $0.92$  & 0.50     \\  
         &$\mu$ stand. $d$ free    & $0.60$ & $1.02$ & -                     & -                     & $6.14$ & $0.92$  & 0.50     \\  
         &$\mu$ simple $d$ fixed   & $0.37$ & $0.63$ & -                     & -                     &  6.64                  & -                       & 0.78     \\ 
         &$\mu$ simple $d$ constr. & $0.44$ & $0.75$ & -                     & -                     & $6.11$ & $0.92$  & 0.55     \\  \vspace{0.048cm}
         &$\mu$ simple $d$ free    & $0.44$ & $0.75$ & -                     & -                     & $6.11$ & $0.92$  & 0.55     \\ 
NGC 2403 &$\mu$ stand. $d$ fixed   & $0.74$ & $1.80$ & -                     & -                     &  3.47                  & -                       & 2.12     \\ 
         &$\mu$ stand. $d$ constr. & $0.62$ & $1.51$ & -                     & -                     & $3.76$ & $1.08$  & 1.43     \\ 
         &$\mu$ stand. $d$ free    & $0.35$ & $0.85$ & -                     & -                     & $4.71$ & $1.46$  & 0.54     \\ 
         &$\mu$ simple $d$ fixed   & $0.53$ & $1.29$ & -                     & -                     &  3.47                  & -                       & 2.29     \\ 
         &$\mu$ simple $d$ constr. & $0.45$ & $1.10$ & -                     & -                     & $3.76$ & $1.08$  & 1.73     \\ \vspace{0.048cm}
         &$\mu$ simple $d$ free    & $0.26$ & $0.63$ & -                     & -                     & $4.69$ & $1.46$  & 0.56     \\ 
NGC 7793 &$\mu$ stand. $d$ fixed   & $0.46$ & $1.48$ & -                     & -                     &  3.91                  & -                       & 6.21     \\ 
         &$\mu$ stand. $d$ constr. & $0.39$ & $1.26$ & -                     & -                     & $4.30$ & $1.10$  & 5.40     \\  
         &$\mu$ stand. $d$ free    & $0.14$ & $0.45$ & -                     & -                     & $7.02$ & $1.80$  & 3.08     \\  
         &$\mu$ simple $d$ fixed   & $0.33$ & $1.06$ & -                     & -                     &  3.91                  & -                       & 5.94     \\ 
         &$\mu$ simple $d$ constr. & $0.28$ & $0.90$ & -                     & -                     & $4.30$ & $1.10$  & 5.17     \\  \vspace{0.048cm}
         &$\mu$ simple $d$ free    & $0.12$ & $0.39$ & -                     & -                     & $6.56$ & $1.68$  & 3.00     \\   
NGC 2976 &$\mu$ stand. $d$ fixed   & $0.33$ & $0.60$ & -                     & -                     &  3.56                  & -                       & 1.73     \\ 
         &$\mu$ stand. $d$ constr. & $0.28$ & $0.51$ & -                     & -                     & $3.90$ & $1.09$  & 1.65     \\ 
         &$\mu$ stand. $d$ free    & $0.04$ & $0.07$ & -                     & -                     & $9.30$ & $2.61$  & 0.70     \\ 
         &$\mu$ simple $d$ fixed   & $0.23$ & $0.42$ & -                     & -                     &  3.56                  & -                       & 1.61     \\ 
         &$\mu$ simple $d$ constr. & $0.20$ & $0.36$ & -                     & -                     & $3.92$ & $1.10$  & 1.50     \\ \vspace{0.048cm}
         &$\mu$ simple $d$ free    & $0.05$ & $0.09$ & -                     & -                     & $7.72$ & $2.17$  & 0.77     \\  
DDO 154  &$\mu$ stand. $d$ fixed   & $0.00$ & $0.00$ & -                     & -                     &  4.30                  & -                       & 4.60     \\ 
         &$\mu$ stand. $d$ constr. & $0.77$ & $2.41$ & -                     & -                     & $3.23$ & $0.75$  & 0.36     \\  
         &$\mu$ stand. $d$ free    & $0.85$ & $2.66$ & -                     & -                     & $3.18$ & $0.74$  & 0.35     \\  
         &$\mu$ simple $d$ fixed   & $0.00$ & $0.00$ & -                     & -                     &  4.30                  & -                       & 6.73     \\ 
         &$\mu$ simple $d$ constr. & $0.50$ & $1.56$ & -                     & -                     & $3.23$ & $0.75$  & 0.41     \\  \vspace{0.048cm}
         &$\mu$ simple $d$ free    & $0.79$ & $2.47$ & -                     & -                     & $3.04$ & $0.71$  & 0.33     \\ 
\end{longtable}

}

\end{document}